\documentclass[conference]{IEEEtran}
\usepackage{amssymb,amsfonts,amsmath, bbm,amstext}
\usepackage{graphicx,times,latexsym,makeidx,xspace,url,comment,subfigure,multirow}
\usepackage{theorem}
\usepackage[compress]{cite}
\usepackage{color}
\usepackage{pstricks}
\usepackage{todonotes}

\usepackage{amsmath}

\newcommand{\secR}[1]{Sec.~\ref{sec:#1}}
\newcommand{\secL}[1]{\label{sec:#1}}
\newcommand{\eqR}[1]{(\ref{eq:#1})}

\newcommand{\figR}[1]{Fig.~\ref{fig:#1}}
\newcommand{\figLC}[2]{
		\caption{#2}
		\label{fig:#1}
}

\newcommand{\reno}[0]{NewReno}
\newcommand{\lp}[0]{LP}
\newcommand{\nice}[0]{NICE}
\newcommand{\ledbat}[0]{LEDBAT}
\newcommand{\btledbat}[0]{\ledbat}

\newcommand{\I}[1]{{\mathbbm{1}_{#1}}}

\begin{document}

\title{Modeling the interdependency of low-priority congestion control and active queue management}
 \author{ 
 \IEEEauthorblockN{YiXi Gong\IEEEauthorrefmark{1}, Dario Rossi \IEEEauthorrefmark{1}, Emilio Leonardi \IEEEauthorrefmark{2}} 
 \IEEEauthorblockA{ \IEEEauthorrefmark{1} Telecom ParisTech, France -- \texttt{firstname.lastname\@enst.fr}}
 \IEEEauthorblockA{\IEEEauthorrefmark{2} Politecnico di Torino, Italy -- \texttt{emilio.leonardi\@polito.it}}
 }

\maketitle

\begin{abstract}
Recently, a negative interplay has been shown to arise when scheduling/AQM techniques and low-priority congestion control protocols are used together: namely, AQM resets the relative level of priority among congestion control protocols. This work explores this issue by (i) studying a fluid model that describes system dynamics of heterogeneous congestion control protocols competing on a bottleneck link governed by AQM and (ii) proposing a system level solution able to reinstate priorities among protocols.
\end{abstract}

\section{Problem statement} 

It comes to no surprise that our daily activities increasingly require ubiquitous Internet access. In a typical day, we call friends with Skype or Gtalk, socialize on Twitter and Facebook, upload pictures to Picasa and Flickr, backup or share data with BitTorrent, Dropbox and Mega, and upload new tunes to GoogleMusic, etc. Moreover, as a side-effect of the proliferation of connected household devices, the need to synchronize data between the numerous appliances arises. As copies of the data are increasingly stored in some datacenters, this translates into frequent upload/downloads to/from the Cloud.
At the same time, the periphery of the Internet infrastructure was designed having in mind that users would mostly be data consumer (as opposite to data producer), as for instance testified by the deployment of Asymmetric Digital Subscriber Line (ADSL) in Europe. While this fact was already challenged by peer-to-peer traffic (P2P), current uploads to the Cloud further clash the infrastructure asymmetry.

This mismatch may lead to ``bufferbloat''~\cite{bufferbloat12cacm}, i.e., very large queuing delays, up to several seconds, experienced by Internet users. While the ``persistently full buffer'' phenomenon is not new~\cite{cheshire96rants}, it has been exacerbated by the ubiquitous presence of significantly large buffers at the access (made relatively cheap by today's technology), that the loss-based congestion control of TCP is apt to fill: as TCP regulates its sending rate  (halving the sending window)  only in occurrence of  \emph{losses}, the buffer is forcibly filled up\footnote{Notice that buffer sizes involved are often small in absolute terms (few KBs), but are relatively large to capacities of the narrow cable, ADSL or WiFi links (few Kbps) in front of these buffers, yielding possibly multi-second queueing delays~\cite{kreibich2010netalyzr}.}. Moreover, while infrastructural solutions to the bufferbloat have been proposed in the literature, such as scheduling (SFQ~\cite{sfq}, DRR~\cite{drr}) and Active Queue Management (RED~\cite{red},  Choke~\cite{choke}), their adoption has been rather limited. The situation has only recently started to change, with worldwide operators implementing scheduling policies in the upstream of the ADSL modem to improve the quality of user experience (e.g., in France, Free implements SFQ since 2005~\cite{free_sfq}), and with new promising AQM techniques under active development such as CoDel~\cite{codel}.

At the same time, recent evolution of Internet application landscape has also seen a proliferations of new applications --or ``apps'', as the Internet is now often accessed through smartphones or portable devices-- that control the flow of traffic to and from the Cloud.  Due to the lack of infrastructural solution to the bufferbloat problem, and since deployment of all-software solution is much easier with respect to infrastructural upgrades, applications have started proposing alternative models to the standard TCP best effort congestion control. Notable examples include Microsoft Background Intelligent Transfer Service (BITS), Picasa background upload option, and BitTorrent low extra delay background transport (LEDBAT)\cite{ledbat_draft}. The latter is especially interesting since BitTorrent still represents a sizeable amount of Internet traffic and, according to Bram Cohen, ``LEDBAT is now the bulk of all BitTorrent traffic, [...] most consumer ISPs have seen the majority of their upload traffic switching to a UDP-based protocol''~\cite{cohen_quora}. The rationale behind the design of LEDBAT is that user ADSL link represent likely the uplink bottleneck, so that congestion is typically \emph{self-induced} by concurrent traffic sessions generated by the same user --such as BitTorrent transfers in parallel with Skype call and other Cloud uploads-- which LEDBAT is designed to avoid. 

Our recent work~\cite{tma13} shows, by means of simulation and experiments, a negative interplay when scheduling/AQM techniques and low-priority congestion control (LPCC) protocols such as LEDBAT are combined: namely, AQM resets the relative level of priority among congestion control protocols. In this work, we first study a fluid model that describes system dynamics when flows adhering to heterogeneous congestion control protocols, such as LEDBAT and TCP, compete on a bottleneck link governed by AQM. Then, we propose a system-level solution able to reinstate priorities among protocols.

The remainder of this papers is organized as follows. \secR{related} overviews closest related work. Fluid model is presented in \secR{model}, and an extensive set of numerical results, gathered on the scenario described in \secR{scenario} are reported in \secR{results} and compared against \verb!ns2! simulations. System design of a feasible solution is then described in \secR{system}, before \secR{conclusion} concludes the paper and outlines our next steps.

\section{Background}\secL{related}

It would be extremely cumbersome to comprehensively retrace over 20 years of Internet research in these few pages. An historical viewpoint is sketched in~\cite{darkbuffers12cacm}: we extend this viewpoint by reporting in \figR{timeline} a timeline of research in scheduling/AQM algorithms and LPCC protocols. The timeline clearly shows a temporal separation of these two research topics, which in our opinion helps understand why the AQM vs. LPCC interaction assessed in this paper was only barely previously exposed. In this section, we overview the related work separately considering (i) the AQM vs. LPCC interaction, (ii) fairness of congestion control protocols, (iii) LEDBAT and other low priority protocols, (iv) fluid modeling  of TCP and AQM.

~\\

\newcommand{\sm}[1]{\begin{scriptsize}#1\end{scriptsize}}
\begin{figure}[!h]
\psset{xunit=1}
\begin{pspicture}(0,0)(9,1)
 \psline{->}(8.5,0)
\uput{12mm}[90](0.25,0){\sm{\emph{LPCC:}}}
\uput{7.5mm}[90](0.25,0){\sm{\emph{AQM:}}}
\uput{3mm}[90](0.25,0){\sm{\emph{[ref]}}}
\uput{3mm}[-90](0.25,0){\sm{\emph{year}}}
%

\psline(1,-0.25)(1,0.25)
\uput{7.5mm}[90](1,0){\sm{SFQ}}
\uput{3mm}[90](1,0){\sm{\cite{sfq}}}
\uput{3mm}[-90](1,0){\sm{1990}}

\psline(2,-0.25)(2,0.25)
\uput{8mm}[90](2,0){\sm{RED}}
\uput{3mm}[90](2,0){\sm{\cite{red}}}
\uput{3mm}[-90](2,0){\sm{1993}}

\psline(3,-0.25)(3,0.25)
\uput{8mm}[90](3,0){\sm{DRR}}
\uput{3mm}[90](3,0){\sm{\cite{drr}}}
\uput{3mm}[-90](3,0){\sm{1995}}

\psline(4,-0.25)(4,0.25)
\uput{8mm}[90](4,0){\sm{Choke}}
\uput{3mm}[90](4,0){\sm{\cite{choke}}}
\uput{3mm}[-90](4,0){\sm{2000}}

\psline(5,-0.25)(5,0.7)
\uput{12mm}[90](5,0){\sm{NICE}}
\uput{8mm}[90](5,0){\sm{\cite{tcp_nice}}}
\uput{3mm}[-90](5,0){\sm{2002}}

\psline(6,-0.25)(6,0.7)
\uput{12mm}[90](6,0){\sm{LP}}
\uput{8mm}[90](6,0){\sm{\cite{tcp_lp}}}
\uput{3mm}[-90](6,0){\sm{2003}}


\psline(7,-0.25)(7,0.7)
\uput{12mm}[90](7,0){\sm{LEDBAT}}
\uput{8mm}[90](7,0){\sm{\cite{ledbat_draft}}}
\uput{3mm}[-90](7,0){\sm{2010}}

\psline(8,-0.25)(8,0.25)
\uput{8mm}[90](8,0){\sm{CoDel}}
\uput{3mm}[90](8,0){\sm{\cite{codel}}}
\uput{3mm}[-90](8,0){\sm{2012}}

\end{pspicture}
\figLC{timeline}{Timeline of AQM and LPCC algorithms.}
\end{figure}

\subsection{Interaction}

To the best of our knowledge, aside our previous work ~\cite{tma13}, only~\cite{itc22nec} mentions AQM and a LPCC (namely, LEDBAT) in the same paper. In one of the tests, the authors experiment with a home gateway that implement some (non-specified) AQM policy other than DropTail. When LEDBAT and TCP are both marked in the same ``background class'', the ``TCP upstream traffic achieves a higher throughput than the LEDBAT flows but significantly lower than'' under DropTail~\cite{itc22nec}. This is known explicitly in the LEDBAT RFC, stating that under AQM is possible that ``LEDBAT reverts to standard TCP behavior, rather than yield to other TCP flows''~\cite{ledbat_draft}.   

In our previous work~\cite{tma13}, we further show that this behavior is general and can arise from the interaction of any scheduling/AQM discipline and LPCC protocol shown in \figR{timeline}, using a twofold methodology including \verb!ns2! simulation and experiments from both controlled testbed and wild Internet. The present work differs from~\cite{tma13} in both its depth and methodology: indeed, we adopt a more narrow but profound scope, selecting LEDBAT and RED as representative examples of the LPCC and scheduling/AQM design space that we then analytically model.

\subsection{Fairness}

Our main focus in this paper concerns fairness of the capacity share among heterogeneous control protocols on a bottleneck governed by AQM. While fairness is a long studied subject, its investigation generally considered rather different settings. First, it has often been tackled in the \emph{intra-protocol} case~\cite{floyd91sigcomm,icccn10,globecom10,tr10}: i.e., heterogeneous settings of a single protocol flavor. For instance, \cite{floyd91sigcomm} studies RTT unfairness of TCP Reno. Similarly, we pointed out the existence of a LEDBAT latecomer unfairness issue~\cite{icccn10} -- that we show to be less  relevant in the case of short lived flows and solve for backlogged 
connections in~\cite{globecom10,tr10}.

Fairness in the \emph{inter-protocol} case, thus closer to ours heterogeneous control protocols settings, has long been studied as well~\cite{ahn95sigcomm,eshete12aintec,eshete23itc}.
Old works especially focused on undesirable side-effect of delay-based congestion control of Vegas, that makes it back off in presence of TCP Reno~\cite{ahn95sigcomm}. 
Even more recent work on the topic studies different issues than ours. Authors of ~\cite{eshete12aintec,eshete23itc} focus on several high-speed variants of TCP: in their case, fairness between the different protocols is thus desirable, while in our settings \emph{unfairness would be desirable} (as it would imply that low-priority property is maintained). 
Complementary to this work, authors in~\cite{eshete23itc} design and analyze an AQM scheme (named AFpFT after Approximate Fairness through Partial Finish Tag), that they show via \verb!ns2! simulations to reinstate fairness in the heterogeneous protocols case~\cite{eshete12aintec}.

\subsection{Low priority}

Protocols such as NICE~\cite{tcp_nice}, LP~\cite{tcp_lp}, 4CP\cite{tcp_4cp} and \cite{tcp_key} share the same low-priority spirit of \ledbat. We carried out a simulation-based comparison of \nice, \lp\ and \ledbat\ in~\cite{lcn10}, showing that \btledbat\ has the lowest level of priority.

Some important differences among the above protocols are worth stressing. \nice~\cite{tcp_nice} extends the delay-based behavior of TCP Vegas with a multiplicative decrease reaction to early congestion (detected when the number of packets experiencing a large delay in an RTT exceeds a given threshold). Differently from \ledbat, that reacts to instantaneous one-way delay (OWD) variations, \nice\ instead react to RTT variations, thus possibly reducing the sending window in one direction due to growing delay in the opposite direction.

\lp~\cite{tcp_lp} enhances the loss-based behavior of \reno\ with an early congestion detection based on the distance of the OWD from a weighted moving average calculated on all observations. In case of congestion, the protocol halves the rate and enters an
inference phase, during which, if further congestion is detected, the congestion window is set to zero and normal \reno\ behavior is restarted. This differs from \ledbat that aims at explicitly bounding the maximum delay introduced in the bottleneck queue, which is particularly important for VoIP, gaming and all other interactive delay-sensitive applications.

\subsection{Fluid modeling}

Other work~\cite{misra00sigcomm,hollot01infocom,liu03sigmetrics,marsan05ton} relate to this as far as its methodology is concerned.
We point out that, since generally a single dominant TCP flavor is modeled~\cite{misra00sigcomm,hollot01infocom} (optionally including unresponsive background traffic~\cite{liu03sigmetrics} or short-lived connections ~\cite{marsan05ton}), the novelty in this context lies in the definition of a fluid model of \emph{heterogeneous} responsive sources, notably including LEDBAT.

As our main innovation is not on the technique per se, but on its application to the study of a particular problem, we resort to classic models for TCP~\cite{liu03sigmetrics} and RED~\cite{misra00sigcomm}, that we extend to incorporate novel popular protocols such as LEDBAT.


\section{Fluid model}\secL{model}

%
%
\begin{figure}[t]
    \begin{center}
        \includegraphics[width=0.45\textwidth]{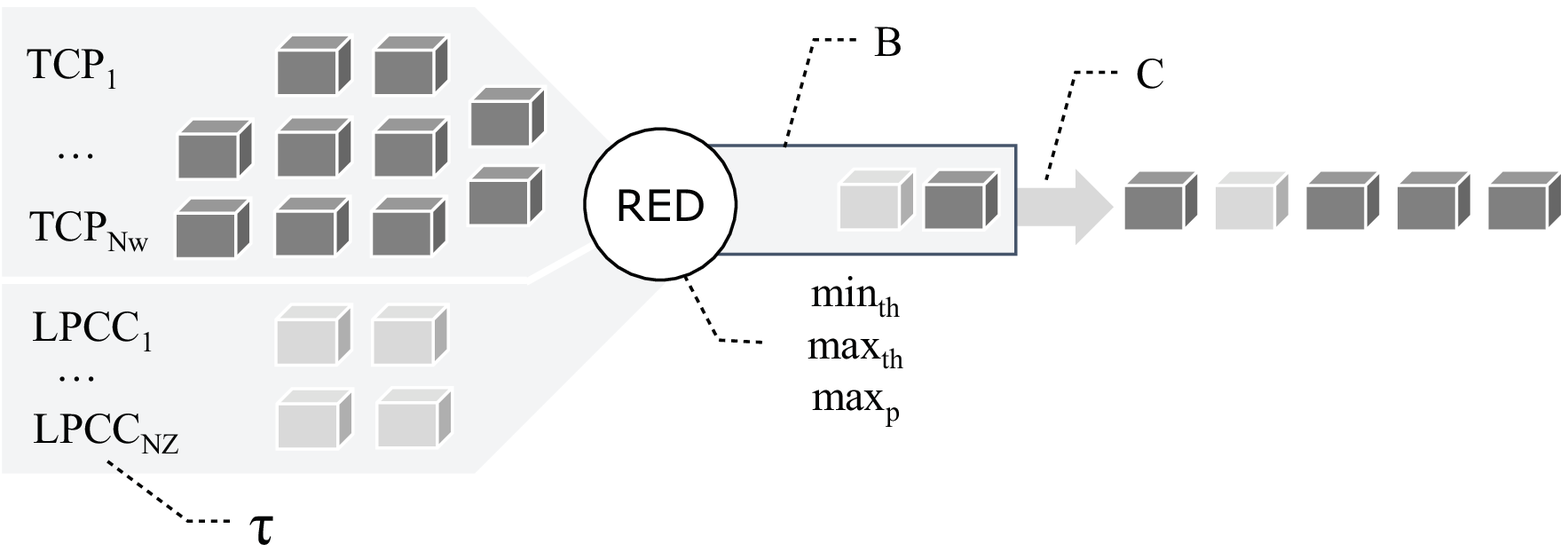}
        \figLC{scenario}{Network scenario}
    \end{center}
\end{figure}
%


We describe the network scenario we model with the help of \figR{scenario}. 
A user device generates a number of long-lived flows competing on the same bottleneck of capacity $C$, with a buffer of size $B$ full size packets. Applications on the device either use a best-effort TCP congestion control, or a lower-than-best-effort LEDBAT control. We denote the TCP and LEDBAT window at time $t$ as $W(t)$ and $Z(t)$ respectively.  

For TCP, we neglect the slow-start phase, which is instead only optional in LEDBAT. As such, we limitedly model the TCP congestion window behavior in congestion avoidance phase. At the reception of the $(n\!+\!1)$-th ack at time $t_{n+1}$, the TCP congestion window is updated as follows:
\begin{align}\label{eq:w}
W(t_{n+1})=\hspace{-1mm}
\begin{cases}
\frac{1}{2} W(t_n) &\text{ if packet loss,}\\
W(t_n)+ \frac{1}{W(t_n)}&\text{ otherwise}
\end{cases}
\end{align}
\noindent As for LEDBAT, it reacts to losses by halving the congestion window as TCP does, but otherwise its congestion window increase is not larger than TCP ramp-up in congestion avoidance, and more precisely is proportional to the distance of the queuing delay $q(t)$ from the delay target $\tau$:
\begin{align}\label{eq:z}
Z(t_{n+1})=\hspace{-1mm}
\begin{cases}
\frac{1}{2} Z(t_n) &\text{ if packet loss,}\\
Z(t_n) + \frac{1}{Z(t_n)}\frac{\tau-q(t_{n})}{\tau}&\text{ otherwise}
\end{cases}
\end{align}
\noindent with the current queuing delay $q(t)$ measured as:
\begin{align}\label{eq:tau}
   &	q(t_n) = D(t_n) - D_{min}\\
   &    D_{min} = \min_n D(t_n)
\end{align}
\noindent where $D(t_n)$ represents the instantaneous one-way delay (OWD) estimate, while the base delay $D_{min}$ is the minimum observed OWD. The rationale is that, over a sufficiently large number of observations $D_{min}$ accurately represents the fixed component of the delay (i.e., propagation delay plus negligible transmission delay, which should be the one found when queues are empty) so that the $D(t_n) - D_{min}$  difference represents the variable component of the delay (i.e., queuing delay plus negligible processing delay). 

Notice that host synchronization over the Internet is known to be hard. As such, it is worth stressing that the OWD estimate $D(t_n)$ is affected by an unknown clock offset between the two endpoints, and is thus of no practical use. Conversely, the offset cancels in the difference operation in \eqR{tau}, which is only affected by clock drift -- that is of much smaller magnitude and furthermore easier to correct~\cite{cohen10iptps}.

From \eqR{z}, we gather that ramp-up is as fast as TCP only whenever the queue is empty \eqR{tau}, i.e., $\lim_{q\rightarrow 0} \frac{\tau-q}{\tau} = 1$ . Furthermore, whenever the queuing delay hits the target $\tau$,  the congestion window settles, i.e., $\lim_{q\rightarrow \tau} \frac{\tau-q}{\tau} = 0$.

\subsection{Ordinary Differential Equation System} 

To analyze the interactions between sources and queue dynamics, we adopt a continuous time fluid approach~\cite{misra00sigcomm,hollot01infocom,liu03sigmetrics,marsan05ton} in which the average dynamics of both sources and queues are described by deterministic Ordinary Differential Equations (ODE).
To write the ODE system, we denote by $W_i(t)$ the instantaneous congestion window at time $t$ for connection $i$ in the fluid system, by $R_i(t)$ the Round Trip Time (RTT) and by $p(t)$ the packet dropping probability at the buffer. We consider the case of $N_W$ TCP and $N_Z$ LEDBAT connections sharing the same bottleneck, where flow-level congestion window evolution the TCP case is adopted from ~\cite{misra00sigcomm}:

\begin{align}
   \frac{dW_i(t)}{dt}= & \frac{1}{R_i(t)} - \frac{W_i(t)W(t-R_i(t))}{2R(t-R_i(t))} p (t-R_i(t)) \label{eq:W} \\
   \frac{dZ_i(t)}{dt}= & \frac{\tau - q(t)/C}{\tau}\frac{1}{R_i(t)} - \frac{Z_i(t)Z(t-R_i(t))}{2R_i(t-R_i(t))}  p(t-R_i(t)) \label{eq:Z} \\
\frac{dq(t)}{dt}= & \left[\sum_{i=1}^{N_W}\frac{W_i(t)}{R_i(t)} +\sum_{i=1}^{N_Z}\frac{Z_i(t)}{R_i(t)}\right](1- p(t))  -C \I{Q(t)\ge 0}  \label{eq:q} 
\end{align}
In accordance with RED specifications the packet dropping probability at the buffer, $p(t)$, is a function $f(\cdot)$ of the estimated average queue size $Q(t)$, with: 
\[
f(Q)= \left\{
  \begin{array}{ll}  
     0  &   Q<  \min_{th} \\
     \max_p \frac{Q-\min_{th}}{\max_{th}-\min_{th} } &  \min_{th} \le Q \le  \max_{th} \\
     1  &    Q > \max_{th}  
 \end{array} \right.
\]
$Q(t)$ is obtained by $q(t)$ through an exponential weighted moving average from samples taken every $\delta$ seconds.
The fluid equation that relates $Q(t)$ to $q(t)$ is given as in~\cite{misra00sigcomm} by:
\begin{equation} 
\frac{d Q (t)}{dt}= - \frac{ log (1- \alpha)}{ \delta}  Q(t)  +   \frac{ log (1- \alpha)}{ \delta } q(t)  \label{eq:Q}
\end{equation}

Notice that in general case, the RTT $R_i(t) = T_{p,i} + q(t)/C$ cannot be considered to be constant since  the component due to queuing delay can be predominant over the propagation delay $q(t)/C > T_{p,i} $ -- which is especially true in case of  bufferbloat due to FIFO buffering. Conversely, in case AQM is used, it could be reasonable to assume the reverse $q(t)/C < T_{p,i}$ to hold.

%
%

\subsection{Equilibrium point} 


The equilibrium point  of the above dynamical system is  given by the stationary (i.e. constant) solution  $(W^*, Z^*, q^*, Q^*)$ of the system of differential equations, (\ref{eq:W}), (\ref{eq:Z}), (\ref{eq:q}) and (\ref{eq:Q}).
In our case, we are going to prove the existence of at most one unique equilibrium point. We remark that the existence of the equilibrium point can be always granted by properly setting the RED parameters.
For the sake of simplicity we consider a homogeneous case, i.e. a case in which all the connections exhibit the same RTT, however we would like to remark that the extension to a more general case is straightforward. 

From (\ref{eq:Q}), (\ref{eq:W}) and (\ref{eq:Z}) respectively, with simple algebra we obtain:
\begin{align}
q^*&=Q^*\\
W^*&=\sqrt{\frac{2}{p^*}}\\
Z^*&=\left\{ \begin{array}{ll}
\sqrt{\frac{2}{p^*}\frac{\tau -  q^*}{\tau }}  & \mbox{if } q^*<\tau\\
   0             & \mbox{if } q^*\ge \tau \\
\end{array}\right.
\end{align}
\noindent where we have ignored the upper/lower clipping effects on the window size in both LEBDAT and TCP.
From (\ref{eq:q}) we obtain that $q^*$ (with $q^* \in [0, \max_{th}]$), has to be the solution of the equation:
\begin{multline}
\left(T_p + q/C\right)\frac{\sqrt{f(q)}}{ 1- {f(q)}}= 
\left\{ \begin{array}{ll}
\frac{ \sqrt{2}N_W+\sqrt{2\frac{\tau- q}{\tau}}N_{Z}}{C}                 & \text{if } q^* < \tau \\
\frac{ \sqrt{2} N_{W}}{C}      &\text{if }  q^*\ge\tau        
\end{array} \right.  \label{eq:eq}  
\end{multline}
Observe that the existence of at most one solution for (\ref{eq:eq}) is granted by the fact that while the expression on the left is weakly increasing with $q$ (it takes the value 0 for $q=0$),  the expression on the right is, instead, weakly decreasing (being strictly positive for any $q>0$). Thus, a unique solution exists iff:
\begin{multline}
\left(T_p +\frac{\max_{th}}{C}\right)\frac{\sqrt{\max_p}}{ 1- \max_p}>\\>
 \left\{  \begin{array}{ll}\frac{\sqrt{2}N_W+\sqrt{2 \frac{\tau-\max_{th} }{\tau}}N_{Z}}{C}  & \text{if }  \tau >\max_{th}\\  
 \frac{\sqrt{2} N_{W}}{C} &  \text{if } \tau\le\max_{th}
\end{array}  \right. \label{eq:excond}
\end{multline}
Observe that by properly setting ${\max_p}$ we can always meet (\ref{eq:excond}), indeed the term on the left tends to infinite as ${\max_p} \to 1$.

As long as RED is configured to keep the queue shorter with respect to the LEDBAT target (i.e., as long as $q^*\le \tau$) a non-perfect prioritization between TCP flows and LEBDAT flows is experienced, indeed LEBDAT flows are still able to grab a non-negligible fraction of the bottleneck bandwidth.  As discussed in \secR{scenario}, this is the most likely case in practice.

\section{Scenarios}\secL{scenario}

\subsection{User applications}
We argue that the most challenging scenario, in terms of matching results gathered via simulation and fluid mode, is the one with few number of flows. This is intuitive since in the case of multiple backlogged connections, statistical multiplexing will smooth out the impact of events, such as TCP retransmission time out, that would otherwise cause discontinuities in the case of few connections. At the same time, we also argue that the most practically relevant scenario is precisely one with a relatively small number of flows. Indeed, since the bottleneck sits at the user access, the number of concurrent connections will be bound, even considering multiple applications/users in the household.

We consider both the Cloud and the P2P cases. In the Cloud case, it is easy to see that a small number of connections will be opened, at any given time, for a specific service. While considering a single user, even the server contacted will evolve over time (e.g., due to load balancing), this likely happens over time-scales that are much larger with respect to the short time-frames that we consider as ``backlogged'' data transfers (i.e., from tens of seconds to minutes) in this paper. Hence, the number of backlogged connections is upper-bounded by the number of Cloud services the user subscribes to, such as DropBox for  data, GoogleMusic for music and Picasa for pictures/videos. Additionally, the number of simultaneous connections also depends on the on/off synchronization pattern toward the Cloud. As users are not continuously generating all kind of data at the same time, it thus reasonable to envision only a moderate number of concurrent backlogged connections per household, some of which may be lower priorities (e.g., pictures) over others (e.g., critical data, backup).

Consider next the P2P case, where it makes sense to consider file-sharing applications such as BitTorrent due to its popularity, and since it introduced LEDBAT in the first place precisely due to the bufferbloat problem.
In BitTorrent, pipelining of piece requests at the application-level can cause multiple chunks to be transmitted consecutively over the same connection at transport-level. Since BitTorrent limits the number of concurrent slots to about\footnote{The limit actually increases with the square root of the uplink capacity} 4 per torrent, the number of concurrent connections will be again small. Moreover, BitTorrent peers periodically evaluate the throughput toward other peers every 10s of seconds, and connections are maintained in case of good end-to-end throughput: coupled to pipelining, this entails that over the tens of seconds to minute timescale, connections can be considered backlogged.

From the above discussion, in the following we will limitedly consider an equal number $N=N_W=N_Z$ of flows, and let the total number of flows vary in  $2N\in[2,10]$ range. Unless otherwise stated, we consider homogeneous RTT delay settings with propagation delay $T_p=50$\,ms (to which we add a jittering component of 1\,ms to avoid synchronization of the congestion window dynamics).
To precisely characterize system equilibrium properties, we will let 
the LEDBAT target $\tau$ vary -- that in the uTorrent implementation of LEDBAT, this can be easily done by tweaking the {\texttt{net.utp\_target\_delay}} settings.

\subsection{Network configuration}

%
%
\begin{figure*}[t]
    \begin{center}
        \includegraphics[angle=-90,width=1\textwidth]{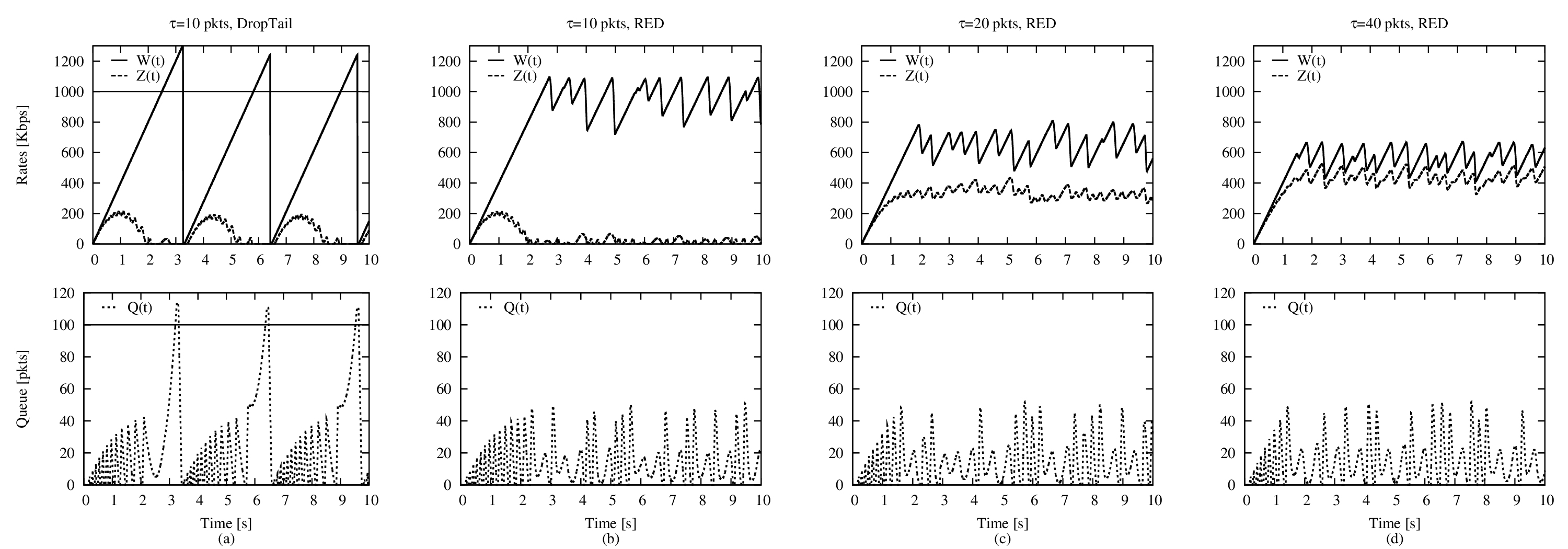}
        \figLC{time}{Reprioritization phenomenon: Time evolution of $W(t)$, $Z(t)$ and $Q(t)$ under DropTail (a) and RED (b,c,d) for different values of $\tau$. }
    \end{center}
\end{figure*}

Without loss of generality, we consider a single access bottleneck and fix the capacity to $C$=1Mbps, typical range for ADSL/Cable access. The bottleneck buffer can accommodate up to $B=100$ packets that, considering $P$=1250\,Bytes sized packets for simplicity, corresponds to a maximum queuing delay of $1000$\,ms. Notice that these values are commonplace nowadays, with modem buffers able to hold up to 4\,seconds worth of traffic~\cite{kreibich2010netalyzr}. 
To precisely characterize system equilibrium properties, we explore variations of the RED settings $\min_{th}, \max_{th}$ (in packets), $\max_p$ settings to cover the full support\footnote{Notice that we do not aim at providing tuning guidelines of RED, which is notoriously difficult~\cite{red_tuning} and scenario dependent~\cite{hollot01infocom}, but rather to provide thorough characterization of the equilibrium.}. For convenience, we may express the target $\tau$ in milliseconds or packets: notice that due to our settings, a packet is worth 10\,ms of queuing delay.

As previously observed, the reprioritization phenomenon vanishes in case $q^* >  \tau$: in other words, when the queue size at the equilibrium exceeds the LEDBAT queuing delay target $\tau$, all LEDBAT flows will by design yield to TCP and the system will behave as~\cite{misra00sigcomm,hollot01infocom}. At the same time, this scenario is unlikely to hold in practice. Consider indeed that end-to-end congestion control protocols such as LEDBAT rely on noisy measures of queuing delay, so that they will not be able to guarantee protocol efficiency when $\tau \rightarrow 0$.
 
Then, notice that for the typical ADSL transmission speed of 500Kbps, the transmission  of a full MTU packet takes about 24\,ms: initial versions of LEDBAT used to set $\tau$ = 25\,ms, i.e., a packet worth of queuing. However, due to practical limitations (including timestamp precision in Windows OS, clock drift of several ppm in off-the-shelf PCs, etc.) this  setting did not allow to fully exploit the link capacity, reason why the target was later increased to $\tau=$100\,ms. While a 100\,ms target may be reasonable for an end-to-end protocol, an AQM may be more precise in measuring the queue size and in adopting more aggressive dropping policy (e.g., $\min_{th} < \max_{th} < \tau$ for RED, or lower packet sojourn time than $\tau$  for CoDel), so that it is reasonable to assume that the target AQM queue size will be $q^* \le  \tau$ in practice.

%

\section{Results}\secL{results}
In this section, we present and discuss numerical results of the ODE describing the system dynamics. Numerical results are gathered either (i) finding roots to the equilibrium equation via bisection method or (ii) integrating the ODE via Runge-Kutta, and are organized as follow. After a description of the scenario (\secR{scenario}), we depict the temporal system evolution to show the reprioritization phenomenon (\secR{reprioritization}) that we will investigate further at the equilibrium (\secR{equilibrium}). We then carry out a sensitivity analysis (\secR{sensitivity}) and discuss local convergence properties (\secR{convergence}) of the model. Finally, we validate a subset of the numerical results against those obtained from \verb!ns2! simulator (\secR{validation}) using our own implementation of LEDBAT, which is available as open source at~\cite{ledbat_code}. Validation is performed on the most challenging (in terms of matching the simulation vs. fluid model results) and relevant  scenarios (in terms of practical relevance).

\subsection{Reprioritization}\secL{reprioritization}

We start by showing the time-evolution of the system equations when $N_W=N_Z=1$ in \figR{time} under either DropTail (a) or RED (b,c,d) disciplines. Top plot shows the $W(t)$, $Z(t)$ and $W(t)+Z(t)$ congestion windows evolution, while queue $Q(t)$ is reported in bottom plots. 

In the DropTail case, we set $\tau=$10 packets and observe the same behavior shown via \verb!ns2! simulation in~\cite{icccn10}: i.e., LEDBAT yields to TCP as expected under DropTail.
In the RED case, we set $\max_{th}=B=100, \min_{th}=10, \max_p=1$ for the sake of illustration and let $\tau$ grow from 10 (b) to 20 (c) and 50 (d) packets.
Notice that in case (b),  RED drastically reduces the queue size and let TCP window fluctuates at about the capacity. Yet, when the target increases in (c) and (d), LEDBAT becomes increasingly aggressive under RED, and competes more fairly against TCP.

To avoid cluttering the pictures, we instead avoid reporting the behavior of LEDBAT for increasing target $\tau$ under DropTail: from the sensitivity simulation-based sensitivity analysis reported in ~\cite{lcn10}, it emerges that
LEDBAT yields to TCP for a large range of $\tau<B$ values, and only whenever $\tau$ approaches (or exceeds) the buffer size $B$ LEDBAT behavior becomes loss-based as TCP.

Shortly, in the following we will refer to this difference in LEDBAT aggressiveness with respect to TCP as a ``reprioritization'' phenomenon induced by RED, which indeed resets the relative level of priorities between LEDBAT and TCP.

\subsection{Equilibrium}\secL{equilibrium}

While it is hard to get closed form solution of the equilibrium point, we can numerically find roots of the ODE equations via the bisection method.
We now characterize the reprioritization as a function of system parameters. For convenience, we define the TCP share ratio as the ratio between $W(t)$ and $Z(t)$: 
\begin{align}
\rho(t) = \frac{W(t)}{W(t)+Z(t)}
\end{align}

\noindent at the equilibrium we have:
\begin{align}
\rho^* = \frac{1}{1 + \sqrt{\frac{\tau- q^*}{\tau}}} \label{eq:rhostar}
\end{align}

%
%
\begin{figure}[t]
    \begin{center}
        \includegraphics[angle=-90,width=0.45\textwidth]{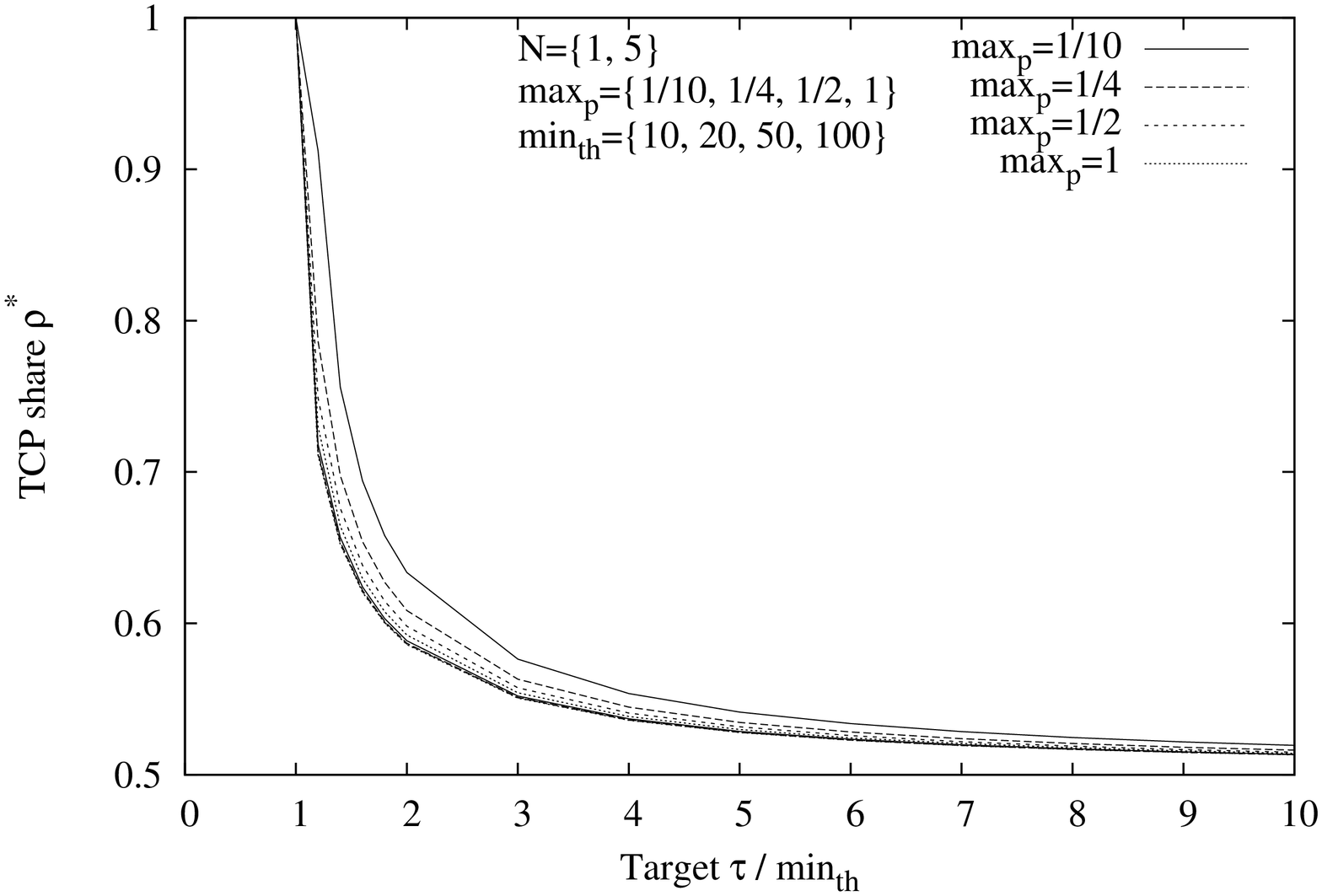}
        \figLC{rho}{Equilibrium analysis of TCP share ratio $\rho^*$ at the equilibrium as a function of $\tau/\min_{th}$ for various flow number $N$, LEDBAT $\tau$, and RED $\min_{th}$, $\max_p$ settings.}
    \end{center}
\end{figure}

Notice that in \figR{time} we have purposely selected settings that show that the system may actually fluctuate around the equilibrium point ($\max_p=1$), though for many settings the equilibrium is actually smoothly reached.
We depict in \figR{rho} the TCP share ratio $\rho^*$ at the equilibrium for varying user scenarios (i.e., number of TCP and LEDBAT flows $N$, LEDBAT target settings $\tau$,  and RED settings $\min_{th}$ and $\max_p$). 

\figR{rho} shows that  under AQM the TCP share exhibits a sharp transition phase as soon as  $\tau$ exceeds $\min_{th}$, quickly dropping with an hyperbolic slope from a monopoly situation ($\rho^* \rightarrow 1 $ for values of $\tau$ close to $q^*$) to a fair share ($\rho^* \approx 0.58 $ for  $\tau = 2q^*$). 
Interestingly, ~\cite{lcn10} shows that in the DropTail case, a sharp transition phase from TCP monopoly to a fair share happens whenever $\tau \rightarrow B$.
This difference is rooted on the fact that RED dropping rates are strictly positive as soon as the queue size exceeds $\min_{th}$, whereas DropTail decisions have to wait until the queue exceeds $B$.

\subsection{Sensitivity}\secL{sensitivity}

From \figR{rho} we also gather that different RED settings yield only minimally affect the reprioritization phenomenon. For completeness, we depict in \figR{sensitivity} the values of the queue $q^*$ (top plot) and the TCP share  $\rho^*$ (bottom plot) at the equilibrium for several RED settings. This time, $q^*$ and $\rho^*$ are reported directly as a function of $\tau$. (i.e., we avoid normalizing over the RED $\min_{th}$ parameter.) 
Trivially, since no dropping happens for $q<\min_{th}$, this parameter plays the biggest role in determining the queue size at the equilibrium. Next comes the load factor, i.e., number of flows insisting on the bottleneck, followed by the maximum dropping probability $\max_{p}$ of the RED profile. 

The impact of the LEDBAT target $\tau$ on the queue size has almost a step-like behavior, that can be explained taking into account that LEDBAT flows activate only when $\tau>q^*$ (or, $\tau>\min_{th}$ given the above remark). 
Recalling the sharp transition phase in LEDBAT aggressiveness as soon as $\tau>\min_{th}$, the impact of LEDBAT flows after activation is to increase the load profile, about as a TCP flow would do.

%
%
\begin{figure}[t]
    \begin{center}
		\includegraphics[angle=-90,width=0.45\textwidth]{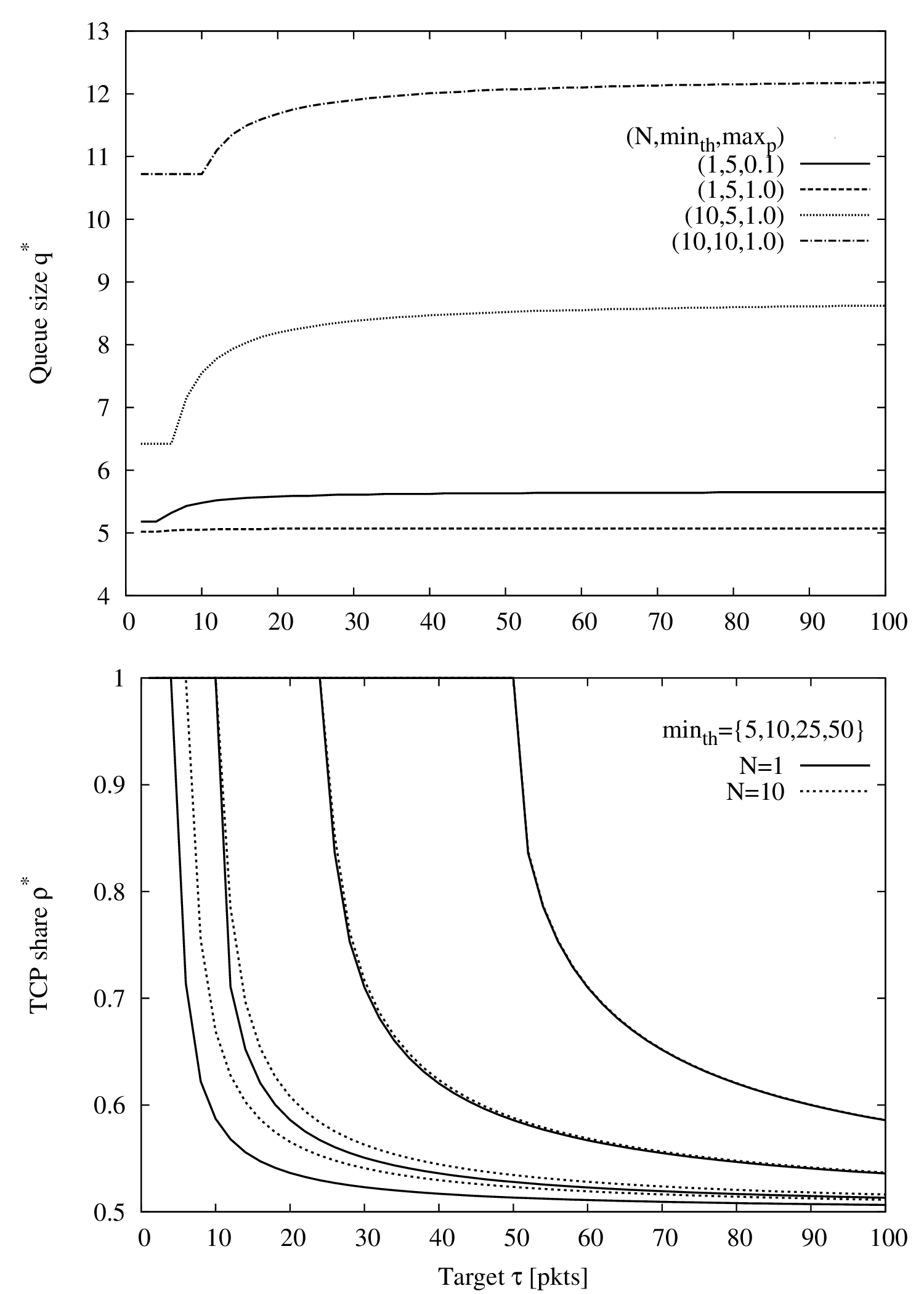}
        \figLC{sensitivity}{Sensitivity analysis of TCP share ratio $\rho^*$ and Queue size $q^*$ at the equilibrium as a function of $\tau$ for various flow number $N$, LEDBAT $\tau$, and RED $\min_{th}, \max_p$ settings.}
    \end{center}
\end{figure}

Bottom plot of \figR{sensitivity} reports similar information to previous \figR{rho}, although lines are now clearly separated for different $\min_{th}$ profiles. We stress that large values of $\min_{th}\ge 50$ could avoid the reprioritization, but at the price of an already sizeable bufferbloat. 

\subsection{Convergence}\secL{convergence}

We now observe evolution of the primitive $W(t),Z(t),q(t)$ variables and of the $\rho(t)$ observable toward the equilibrium $W^*,Z^*,q^*,\rho^*$. Some examples of trajectories are shown in \figR{convergence}.
In more details, top plots report the case for initial conditions $W(0)=Z(0)=q(0)=0$, while bottom plots report the trajectories of 100 random initial conditions. For convenience, we express the relative error with respect to the equilibrium, so that we can superpose multiple trajectories on the same graph.

Top-left plot shows the relative distance of $(W(t),\rho(t))$ from the equilibrium ($W^*,q^*$), while top-right plot shows $(W(t),q(t))$, for two different scenarios. Especially, it can be seen that after an initial oscillation, the queue converges to the equilibrium (also reflected in the breakdown), while convergence is smoother for the other variables.

Bottom-left plot considers 100 random initial conditions and focuses on the initial phase ($t<t'$)  of the system evolution  shown in the top-left counterpart. Similarly, the bottom-right plot considers 100 random initial conditions but focuses on later times when the system is about to reach convergence ($t>t''$).

While further steps are necessary to prove the local system stability (e.g., studying a linearized version of the system at the equilibrium, which is part of our ongoing work), this simple visual inspection has already provided useful insights about the convergence of the equilibrium point for different initial conditions and scenarios.

\subsection{Validation}\secL{validation}

We confirm the validity of the model by contrasting in \figR{validation} the value $\rho^*$ of the TCP share at the equilibrium against simulation results obtained via our own LEDBAT \verb!ns2! implementation~\cite{ledbat_code}. 
We point out that we have already extensively analyzed the reprioritization phenomena via both experiments and simulations~\cite{tma13}, making the \verb!ns2! scripts available at~\cite{ledbat_interaction} to reproduce the phenomenon. Hence, our main aim here is not to provide a coverage of those results, but rather to validate the most representative instance of our results -- which is clearly represented by the TCP share ratio that precisely quantifies the reprioritization.

As we have previously seen, $\min_{th}$ has by far the biggest role in determining the TCP share curve, followed by the number of flows in the bottleneck and by $\max_{p}$ at last. At the same time, while the traffic scenario depends on the user and is a free parameter, from the discussion in \secR{scenario} we do not consider $\min_{th}$ as a free parameter, while $\max_{p}$ is less interesting to study due to its more limited impact.

%

%
%
\begin{figure*}[t]
    \begin{center}
\includegraphics[angle=-90,width=1\textwidth]{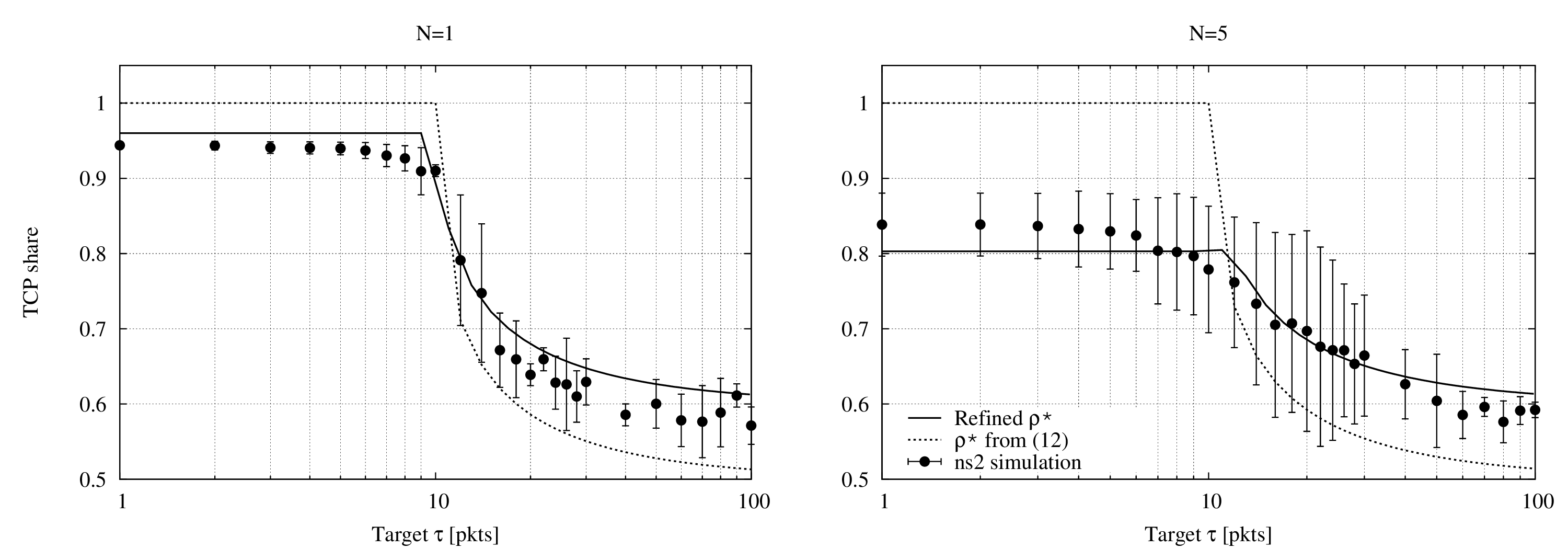}
        \figLC{validation}{Simulation validation of TCP share ratio $\rho^*$ at the equilibrium as a function of $\tau$ for various traffic scenarios   $N_W=N_Z=\{1,5\}$.}
    \end{center}
\end{figure*}

Hence, we fix $\min_{th}=10$, $\max_{th}=B=100$,  $\max_{p}=0.1$ and consider two traffic scenarios $N_W=N_Z=\{1,5\}$. \figR{validation} contrasts average simulation results (solid point, with standard deviation bars over multiple runs) against the equilibrium \eqR{eq} previously discussed (dotted line) and a slightly more accurate version (solid black line) that compensates two simplifications of the fluid model that we discuss next. 
Notice indeed that \eqR{eq} captures reasonably well the essence of the reprioritization phenomenon. Still, two quantitative discrepancies arise.

First, it can be seen that for values of $\tau>\min_{th}$ the model underestimates the TCP share. This results from a known  problem of the TCP model presented in ~\cite{liu03sigmetrics} that  this work extends: i.e.,~\cite{liu03sigmetrics} is known to underestimate TCP congestion window with respect to simulation, which can be easily compensated by taking into account a multiplicative decrease factor of 1.5 (instead of 2) as in~\cite{liu03sigmetrics}.
The refined equilibrium takes into account this correction, and is significantly more accurate when $\tau>\min_{th}$.

Second, recall that when $\tau < q^*$, the model degenerates into a simpler one in which only TCP flows compete on the bottleneck, hence $\rho^*=1$. In practice however, we know that LEDBAT will keep sending a minimum of 1 packet per RTT: this is done to continuously measure the queuing delay, at very low frequency and intrusiveness. LEDBAT does this in order to promptly react to queuing delay reduction and effectively utilize the spare capacity as soon as the link becomes free again. Hence, in case $\tau < q^*$, a refined estimation could upper bound $\rho^*$ by reducing the capacity available for TCP proportionally to the number of LEDBAT flows, i.e.,  

\[\rho^*< 1 - \frac{N}{\frac{C Tp}{P} + q^*} \]

The refined equilibrium takes into account also this second correction, and is significantly more accurate with respect to \eqR{eq} when  $\tau<\min_{th}$. Yet, we argue that such low level of detail can be better captured with \verb!ns2! simulations, and that quantifying the \emph{exact} level of reprioritization is less relevant for practical purposes -- i.e., as users will likely be interested in knowing whether their non-critical bulky transfers are indeed lower-priority with respect to critical continuous backups, or if they compete on a roughly equal basis.


%

%
%
\begin{figure}[t]
    \begin{center}
        \includegraphics[angle=-90,width=0.45\textwidth]{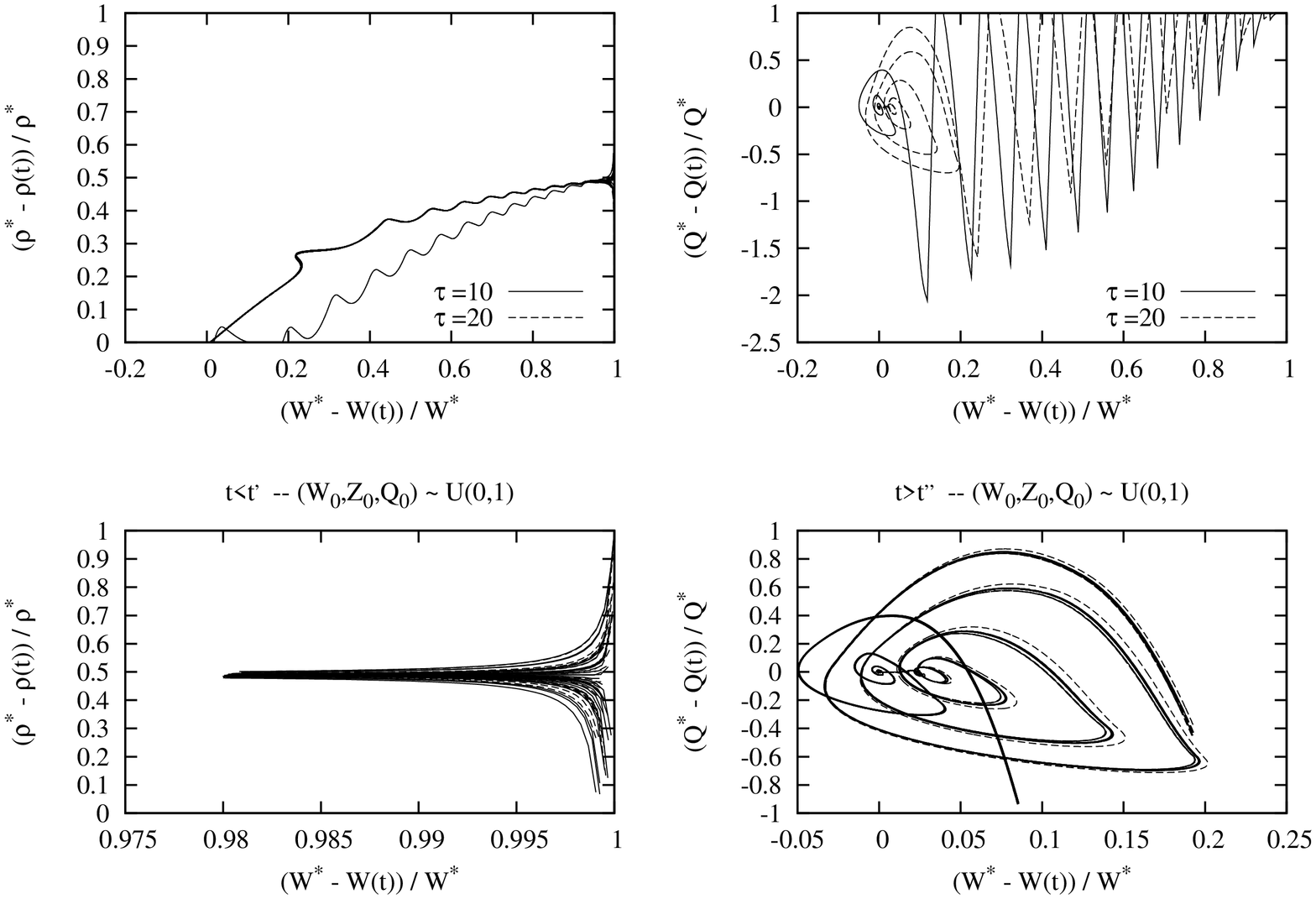}
        \figLC{convergence}{Convergence to the equilibrium.}
    \end{center}
\end{figure}

\section{System-level solution}\secL{system}

Recent evolutions on the Internet applications and infrastructure seem to suggest that AQM and Low priority congestion control (LPCC) protocols will have to coexist: indeed, popular applications are developing delay-based congestion control protocols such as BitTorrent/LEDBAT on the one hand, operators are starting to deploy AQM/scheduling on the user access uplink on the other hand. As such, it is imperative to find solutions to the negative AQM/LPCC interplay we have shown in this paper. While a general solution is hard to find, as testified by the current standpoint after over 20 years of research, a patch to this specific problem may be within reach.

Some might argue that small buffers would be enough to solve bufferbloat altogether. Yet, there are several reasons why this simple solution is not sufficient. First, in presence of too small buffers, it would be difficult for TCP and other congestion control to fully saturate the capacity, causing an undesirable efficiency loss. Second, deciding a buffer size is a matter of concern per se: consider indeed WiFi links, where the capacity may fluctuates widely over time, so that no single buffer size can at the same time (i) be large enough to support TCP congestion control and (ii) rule out bufferbloat in a fast-to-slow transition from 54Mbps to 2Mbps. Finally, jeopardization of relative priorities are not solved by small buffers~\cite{tma13}.

An ideal solution should achieve two goals: (i) meet quality of service constraints while (ii) respecting relative levels of priorities among protocols. Quality of service constraints clearly translate into upper-bounding the queuing delay, that we know is used by protocols to enforce their relative priorities. 
Since even a single TCP flow may bufferbloat the others, the solution \emph{needs AQM}, as otherwise the quality of service constraints would be violated. At the same time, to avoid the LPCC reprioritization phenomenon, we argue that classification capabilities will be needed in AQM to account for flows' \emph{explicitly advertised} level of priority. 

Although in the more general case classification has failed to be adopted (IP TOS field, DiffServ, etc.), and the ability to claim \emph{higher} priority could be easily gamed, in a hybrid AQM vs. LPCC world it makes sense for flows to claim a \emph{lower} priority. We believe that this subtle difference can make an important practical difference in terms of deplorability.

A simple way would be to let application exploit IP TOS field. While the overloading of the IP TOS can be troublesome within an operator network, this is not an issue in the home. Indeed, the usefulness of the IP TOS is not end-to-end but merely meant as a low-priority signal to the box in the user home. Hence, IP TOS could be leveraged by the ISP CPE in the user home to apply differential treatment to best-effort and low-priority traffic (e.g., different AQM loss profiles, different scheduling weights), after which the end-user IP TOS value is no longer useful and can be rewritten by the CPE  (or at the DSLAM, or BRAS, etc.) in the network of an operator using DiffServ if needed.

We further stress that the firmware governing home-routers and WiFi APs is generally based on some variants of the Linux kernel, possibly open-source as in the OpenWrt or CeroWrt cases. We point out that the above solution is therefore already implementable without any additional development effort -- e.g., using strict priority queuing or shaping. In the Linux traffic control (\texttt{tc}) suite, this can be achieved with the \texttt{PRIO} queuing discipline (\texttt{qdisc}) that implements non-shaping container for a configurable number of classes which are dequeued in order. This first solution allows for easy prioritization of traffic, where lower classes are only able to send if higher ones have no packets available. A second solution offered by Linux \texttt{tc} is represented by the \texttt{CBQ qdisc} that offers shaping and finer-grained prioritization capabilities. 

%
%
%
%
%

\section{Conclusions and future work}\secL{conclusion}

This work models an interdependency phenomenon between heterogeneous congestion control protocols and active queue management.
Specifically, in case a low-priority congestion control protocol (e.g., LEDBAT) competes against TCP on a bottleneck link governed by AQM (e.g., RED), the relative level of priority of the congestion control protocols is reset, and the protocols compete on a roughly equal basis.

We model the problem as a system of Ordinary Differential Equations that we solve numerically and validate against simulation results. Our main results is that the TCP share at the equilibrium equals $\rho^* = \big(1 + \sqrt{(\tau- q^*)/\tau}\big)^{-1}$, where $\tau$ is LEDBAT target queuing delay and $q^*$ is the length of the queue at the equilibrium determined by RED settings.

By reason of increasing deployment of both low-priority congestion control and AQM techniques, the problem may be of significant practical relevance. As we believe that it may be desirable for end-users (or end-user applications) to autonomously and coarsely set their relative level of priorities, we have proposed simple yet effective system-level design and practices that can solve the issue we have characterized in this paper.

Our future work moves along two main paths. On the one hand, we aim at refining our investigation by means of a control theoretic analysis of the properties of the linearized system, to e.g., prove local stability of the equilibrium.
On the other hand, as we know by ~\cite{tma13} the problem to hold in general for several AQM and LPCCs, another line of work goes in the direction of extending the model in both the AQM (e.g., CoDel~\cite{codel}) and low-priority congestion control (e.g., \nice~\cite{tcp_nice}) directions. 

\section*{Acknowledgments}
This work was carried out at LINCS \url{http://www.lincs.fr}.  The research leading to these results has received funding from the European Union under the FP7 Grant Agreement n. 318627 (Integrated Project "mPlane").

\bibliographystyle{plain}

\begin{thebibliography}{10}

\bibitem{ledbat_code}
\url{http://www.enst.fr/~drossi/ledbat}.

\bibitem{ledbat_interaction}
\url{http://www.enst.fr/~drossi/dataset/ledbat+aqm}.

\bibitem{free_sfq}
Aduf - historique firmware, 01/07/2005.
\newblock \url{http://88.191.250.12/viewtopic.php?t=164746&view=previous}.

\bibitem{ahn95sigcomm}
J.~Ahn, P.~Danzig, Z.~Liu, and L.~Yan.
\newblock Evaluation of tcp vegas: emulation and experiment.
\newblock In {\em ACM SIGCOMM Computer Communication Review}, volume~25, pages
  185--195. ACM, 1995.

\bibitem{cohen_quora}
B.Cohen.
\newblock How has bittorrent as a protocol evolved over time.
\newblock \url{http://www.quora.com/BitTorrent-protocol-company}.

\bibitem{lcn10}
G.~Carofiglio, L.~Muscariello, D.~Rossi, and C.~Testa.
\newblock {A hands-on Assessment of Transport Protocols with Lower than Best
  Effort Priority}.
\newblock In {\em IEEE LCN}, 2010.

\bibitem{globecom10}
G.~Carofiglio, L.~Muscariello, D.~Rossi, and S.~Valenti.
\newblock {The quest for LEDBAT fairness}.
\newblock In {\em Globecom}, 2010.

\bibitem{bufferbloat12cacm}
V.~Cerf, V.~Jacobson, N.~Weaver, and J.~Gettys.
\newblock Bufferbloat: what's wrong with the internet?
\newblock {\em Communications of the ACM}, 55(2):40--47, 2012.

\bibitem{cheshire96rants}
S.~Cheshire.
\newblock It's the latency, stupid!
\newblock \url{http://rescomp.stanford.edu/~cheshire/rants/Latency.html}, 1996.

\bibitem{red_tuning}
M.~Christiansen, K.~Jeffay, D.~Ott, and F.D. Smith.
\newblock Tuning red for web traffic.
\newblock In {\em ACM SIGCOMM CCR}, volume~30, pages 139--150, 2000.

\bibitem{cohen10iptps}
B.~Cohen and A.~Norberg.
\newblock {Correcting for clock drift in uTP and LEDBAT}.
\newblock In {\em {9th USENIX International Workshop on Peer-to-Peer Systems
  (IPTPS'10)}}, 2010.

\bibitem{eshete23itc}
A.~Eshete and Y.~Jiang.
\newblock Approximate fairness through limited flow list.
\newblock In {\em Proceedings of the 23rd International Teletraffic Congress
  (ITC23)}, pages 198--205, 2011.

\bibitem{eshete12aintec}
A.~Eshete, Y.~Jiang, and L.~Landmark.
\newblock Fairness among high speed and traditional tcp under different queue
  management mechanisms.
\newblock In {\em Proceedings of the ACM Asian Internet Engineeering Conference
  (AINTEC)}, pages 39--46. ACM, 2012.

\bibitem{red}
S.~Floyd and V.~Jacobson.
\newblock Random early detection gateways for congestion avoidance.
\newblock {\em IEEE/ACM Transactions on Networking}, 1(4):397--413, 1993.

\bibitem{floyd91sigcomm}
Sally Floyd.
\newblock Connections with multiple congested gateways in packet-switched
  networks part 1: one-way traffic.
\newblock {\em ACM SIGCOMM Computer Communication Review}, 21(5):30--47, 1991.

\bibitem{tr10}
D.~Rossi C. Testa S.~Valenti G.~Carofiglio, L.~Muscariello.
\newblock Rethinking low extra delay backtround transport protocols.
\newblock {\em Elsevier Computer Networks, to appear}.

\bibitem{darkbuffers12cacm}
J.~Gettys and K.~Nichols.
\newblock Bufferbloat: dark buffers in the internet.
\newblock {\em Communications of the ACM}, 55(1):57--65, 2012.

\bibitem{tma13}
Y.~Gong, D.~Rossi, C.~Testa, S.~Valenti, and D.~Taht.
\newblock Fighting the bufferbloat: on the coexistence of aqm and low priority
  congestion control.
\newblock In {\em IEEE INFOCOM Workshop on Traffic Measurement and Analysis
  (TMA'13)}, Turin, Italy, April 14-19 2013.

\bibitem{hollot01infocom}
CV~Hollot, V.~Misra, D.~Towsley, and W.~Gong.
\newblock A control theoretic analysis of red.
\newblock In {\em IEEE INFOCOM'01}, volume~3, pages 1510--1519, 2001.

\bibitem{tcp_key}
P.~Key, L.~Massouli{\'e}, and B.~Wang.
\newblock {Emulating low-priority transport at the application layer: a
  background transfer service}.
\newblock In {\em ACM SIGMETRICS}, New York City, NY, June 2004.

\bibitem{kreibich2010netalyzr}
C.~Kreibich, N.~Weaver, B.~Nechaev, and V.~Paxson.
\newblock Netalyzr: Illuminating the edge network.
\newblock In {\em ACM SIGCOMM Internet Measurement Conference (IMC'10)}, pages
  246--259, 2010.

\bibitem{tcp_lp}
A.~Kuzmanovic and E.W. Knightly.
\newblock {TCP-LP}: A distributed algorithm for low priority data transfer.
\newblock In {\em IEEE INFOCOM}, 2003.

\bibitem{tcp_4cp}
S.~Liu, M.~Vojnovic, and D.~Gunawardena.
\newblock {4cp: Competitive and considerate congestion control protocol}.
\newblock In {\em ACM SIGCOMM}, September 2006.

\bibitem{liu03sigmetrics}
Y.~Liu, F.~Lo Presti, V.~Misra, D.~Towsley, and Y.~Gu.
\newblock Fluid models and solutions for large-scale ip networks.
\newblock In {\em ACM SIGMETRICS}, volume~31, pages 91--101. ACM, 2003.

\bibitem{marsan05ton}
M~Ajmone Marsan, Michele Garetto, Paolo Giaccone, Emilio Leonardi, Enrico
  Schiattarella, Tarello, and A.
\newblock Using partial differential equations to model tcp mice and elephants
  in large ip networks.
\newblock {\em IEEE/ACM Transactions on Networking,}, 13(6):1289--1301, 2005.

\bibitem{sfq}
P.E. McKenney.
\newblock Stochastic fairness queueing.
\newblock In {\em IEEE INFOCOM}, 1990.

\bibitem{misra00sigcomm}
V.~Misra, W.~Gong, and D.~Towsley.
\newblock Fluid-based analysis of a network of aqm routers supporting tcp flows
  with an application to red.
\newblock {\em ACM SIGCOMM}, 30(4):151--160, 2000.

\bibitem{codel}
K.~Nichols and V.~Jacobson.
\newblock Controlling queue delay.
\newblock {\em Commun. ACM}, 55(7):42--50, July 2012.

\bibitem{choke}
R.~Pan, B.~Prabhakar, and K.~Psounis.
\newblock Choke - a stateless active queue management scheme for approximating
  fair bandwidth allocation.
\newblock In {\em IEEE INFOCOM}, 2000.

\bibitem{icccn10}
D.~Rossi, C.~Testa, S.~Valenti, and L.~Muscariello.
\newblock {LEDBAT: the new BitTorrent congestion control protocol}.
\newblock In {\em IEEE ICCCN}, 2010.

\bibitem{itc22nec}
J.~Schneider, J.~Wagner, R.~Winter, and H.J.Kolbe.
\newblock Out of my way -- evaluating low extra delay background transport in
  an {ADSL} access network.
\newblock In {\em ITC22}, 2010.

\bibitem{ledbat_draft}
S~Shalunov.
\newblock {Low Extra Delay Background Transport (LEDBAT)}.
\newblock {IETF Draft}, March 2010.

\bibitem{drr}
M.~Shreedhar and George Varghese.
\newblock Efficient fair queueing using deficit round robin.
\newblock In {\em ACM SIGCOMM}, 1995.

\bibitem{tcp_nice}
A.~Venkataramani, R.~Kokku, and M.~Dahlin.
\newblock {TCP Nice: A mechanism for background transfers}.
\newblock In {\em USENIX OSDI}, 2002.

\end{thebibliography}

\end{document}